\documentstyle[12pt,epsfig]{article}
\baselineskip 2 cm\date{}
\title{ Sensitivity of single crystals to the circular polarization of 
high-energy $\gamma$ - quanta} 
\author{V.A.Maisheev \thanks{E-mail maisheev@mx.ihep.su}
 \\{\it Institute for High Energy Physics, 142284, Protvino, Russia }}
\begin{document}\maketitle\def\arcctg{\mathop{\rm arccot}
\nolimits} \def\ch{\mathop{\rm ch}\nolimits}\def\sh{\mathop{\rm sh}\nolimits} 
\def\Im{\mathop{\rm Im}\nolimits}\def\Re{\mathop{\rm Re}\nolimits}
\def\sign{\mathop{\rm sign}\nolimits}
{\begin{abstract}
It is shown that single crystals are sensitive 
to the initial circular polarization of  $\gamma$-quanta with energies 
in tens GeV and more. 
The possibility of measurement of $\gamma$-beam polarization is discussed.
The obtained results may be useful for creation of polarimeters for high 
energy beams of $\gamma$-quanta.


\end{abstract}

\section{ Introduction} The birefringence of $\gamma$-quanta with energies 
$ >1 $ GeV propagating in single crystals was predicted in \cite{C}. 
The main process by which $\gamma$-quanta are absorbed in single crystals 
is the electron-positron pair production \cite{TM,U}. 
The cross section of the process depends on the direction of linear 
polarization of the $\gamma$-quanta relative to the crystallographic planes. 
As a result of interaction with the electric field of the single crystal, 
a monochromatic, linearly polarized beam of $\gamma$-quanta comprises 
two electromagnetic waves with different refractive indices, so that 
linear polarization is transformed into circular polarization or vice versa. 
This polarization phenomenon should be observed for symmetric orientations 
of single crystals with respect to the direction of motion 
of  $\gamma$-quanta.
Semilar effects for the production or analysis of polarized $\gamma$-quanta 
have been discussed in \cite{KK}-\cite{BKS}.  

   The anisotropic medium is determined as medium, whose optical properties 
can be described using a symmetric permittivity tensor\cite{AG}. 
The single crystals serve as examples of the similar medium for 
$\gamma$- quanta with energies in tens GeV and more.
The general case of the propagation of high energy $\gamma$-quanta in
the anisotropic medium was considered in \cite{MV1}. This process
was investigated in detail for propagation in a dichromatic laser wave,
which is the simplest sample of the anisotropic medium  of general type.  
(The dichromatic wave is a superposition of the two linearly polarized
laser waves with different frequencies moving in the same direction and, 
broadly speaking, nonzero angle between directions of polarization of
these waves.) The cited paper shows that in the general case the unpolarized 
beam of $\gamma$-quanta obtains some degree of circular polarization
(in contrast to the case in \cite{C}). 

In this paper we examine the propagation of high energy $\gamma$-quanta
in single crystals oriented in some regions of coherent $e^{\pm}$-pair 
production \cite{TM,U}. We show that single crystals are sensitive to 
the circular polarization of propagating $\gamma$-quanta. 

\section{Propagation of $\gamma$-quanta in single crystals.}
The permittivity tensor in  single crystals oriented in the regions
corresponding to coherent $e^{\pm}$-pair production was obtained 
with the help of dispersion relations in \cite{MMF}. 
The components of this tensor are  complex values and have the
following form: $\varepsilon_{ij}= \varepsilon'_{ij}+ i\varepsilon''_{ij},
(i,\, j =1,2)$ (the process is determined by the transverse part of the 
tensor).
It is the symmetrical tensor (i.e., $\varepsilon_{12}= \varepsilon_{21}$) with
components depending on the energy of $\gamma$-quanta  and   two orientation 
angles of the single crystal with respect to the direction of
$\gamma$-beam motion. 

It is well-known that in the general case the symmetric complex tensor is not
reduced to principal axes (i.e., there does not exist a coordinate system  
in which the tensors $\varepsilon'_{ij}$ and $\varepsilon''_{ij}$ are
both diagonal). It is really, this situation is realized in single crystals 
at orientations when some "strong" planes act together on
propagating $\gamma$-quanta \cite{MMF}. Obviously,  these orientations
take a place, when a beam of $\gamma$-quanta move under relatively small
angles with respect to one of crystallographic axes. 

Knowing the permittivity tensor $\varepsilon_{ij}$, one can find
the refractive indices of $\gamma$-quanta \cite{MMF}
\begin{equation} 
\tilde n^2 
=(\varepsilon_{11}+\varepsilon_{22})/2 \pm 
\sqrt{(\varepsilon_{11}-\varepsilon_{22})^2/4 + 
\varepsilon_{12}\varepsilon_{21} }\, ,\label{17}
\end{equation}
Consequently, a beam of $\gamma$-quanta propagates in the single crystals
as a superposition of two electromagnetic waves, which have different 
refractive indices in general. 
One can describe the polarization state either (of the two) wave by the use of 
the Stokes parameters $X_i, Y_i,\,(i=1,3)$ ($X_i$-values correspond to one wave
 and $Y_i$ to another).
These parameters are determined by the following relations\cite{MV1}
\begin{eqnarray}
X_1={\kappa + \kappa^{*} \over 1+\kappa \kappa^{*}} \, ,
  \\
X_2= {i(\kappa - \kappa^{*}) \over 1+ \kappa \kappa^{*}} \, , \label{35} \\
X_3= {\kappa \kappa^{*} -1 \over 1 +\kappa \kappa^{*}} \, , \\
Y_1=-X_1, Y_2=X_2, Y_3= -X_3, X_1^2+X_2^2+X_3^2=1,
\end{eqnarray}
where $\kappa$ is the ratio of components of electric induction
vector ${\bf{D}}(D_1,D_2,0)$ in the coordinate system, in which one axis
is parallel to  the wave vector of $\gamma$-quanta:
\begin{eqnarray}{D_1\over D_2}=\kappa={{ \tilde n^{2}-\varepsilon_{22}} 
\over \varepsilon_{21}} 
\end{eqnarray}
We call attention to relation: $X_2=Y_2$ (i.e., waves have the same value of  
circular polarization). These waves, described here, are the eigenfunctions
of problem, and they named as normal electromagnetic waves \cite{AG}. 
 In general these waves are elliptically polarized at propagation of 
 $\gamma$-quanta in single crystals \cite{MMF}.

Representing the beam of $\gamma$-quanta by a superposition of two normal 
electromagnetic waves with previously determined refractive
indices and polarization characteristic, we obtain  relations describing 
the variation of the intensity and Stokes parameters of the $\gamma$-quanta 
propagating in the single crystal \cite{MV1}:
\begin{eqnarray} 
J_{\gamma}(x)= J_a(x)+J_b(x)+2J_c(x) \, , 
\\ \xi_1(x)= (X_1 J_a(x) + Y_1 J_b(x)+ p_1 J_d(x))/J_{\gamma}(x)\, ,
\\ \xi_2(x) =(X_2 J_a(x) + Y_2 J_b(x) + p_2 J_c(x))/J_{\gamma}(x) \, , 
\\ \xi_3(x)=(X_3 J_a(x) + Y_3 J_b(x) +p_3 J_d(x))/J_{\gamma}(x) 
\\ p_1=- {2X_3 \over X_2}, \, \, \, p_2={2\over X_2}, \, \, \, 
p_3={2X_1\over X_2},
\end{eqnarray}
where  $J_{\gamma}(x), \xi_1(x), \xi_2(x), \xi_3(x)$  are the intensity 
and Stokes parameters of $\gamma$-quanta on the single crystal thickness 
equal to x. Besides, we assume that $J_{\gamma}(0)=1$. 
The partial intensities $J_i(x), \,(i=a, b, c, d)$ have the following form: 
\begin{eqnarray}J_a(x)=J_a(0)\exp(-2\Im(\tilde n_1)\omega x/c) \,
 ,\qquad \qquad \qquad  
\\ J_b(x)=J_b(0)\exp(-2\Im(\tilde n_2)\omega x/c) \, ,\qquad \qquad \qquad  \\
J_c(x)=\exp(-\Im(\tilde n_1 +\tilde n_2 ) \omega x/c)\{J_c(0)\cos(\Re(\tilde 
n_1 -\tilde n_2)\omega x/c) + \\ 
+  J_d(0) \sin(\Re(\tilde n_1 - \tilde n_2) \omega x/c)\} \qquad \qquad 
\qquad \,, \nonumber \\
J_d(x)=-\exp(-\Im(\tilde n_1+ \tilde n_2) \omega x/c)\{J_c(0) \sin(\Re
(\tilde n_1 -\tilde n_2) \omega x/c) -  \\
-J_d(0)\cos(\Re(\tilde n_1- \tilde n_2)\omega x/c)\} \qquad \qquad \qquad \,, 
\nonumber
\end{eqnarray} 
where $\omega$ is the frequency of $\gamma$-quanta and $c$ 
is the speed of light.
One can understand the physical meaning of $J_i$-values, 
if we write the matrix
${\bf{DD^*}}$ in the component-wise form, 
where ${\bf{D}}={\bf{D}_1}+{\bf{D}_2}$ 
is a superposition  of the normal waves.
The initial partial intensities are defined from relations:
\begin{eqnarray}
J_a(0)= {{1+X_1\xi_1(0) -X_2\xi_2(0)+ X_3\xi_3(0)}\over { 2(X_1^2+X_3^2)}}, \\
J_b(0)= {{1-X_1\xi_1(0) -X_2\xi_2(0)- X_3\xi_3(0)}\over { 2(X_1^2+X_3^2)}},\\
J_c(0)={{X_2\xi_2(0)-X_2^2}\over{2(X_1^2+X_3^2)}}, \\
J_d(0)={{X_2(X_1\xi_3(0)-X_3\xi_1(0))}\over{2(X_1^2+X_3^2)}}.
\end{eqnarray}

For small thickness one can obtain the following relation
\begin{eqnarray}
J_{\gamma}(x) = 1- N\sigma_\gamma(\xi_1(0),\xi_3(0)) x \, ,
\end{eqnarray}
where $\sigma_\gamma$ is the total cross section \cite{BKS}
 of $e^{\pm}$-pair production
by $\gamma$-quanta with the initial  $\xi_1(0), \xi_3(0)$-Stokes parameters
and  N is the number of atoms per unit volume. Notice that the dependence
on $\xi_2(0)$  appears only at $x^2$-terms (and higher) in the expansion
(if $X_2 \ne 0$).   

One can see that the initially unpolarized beam of $\gamma$-quanta 
($\xi_i(0)=0,\,i=1-3$) obtains some degree of linear and circular polarization
at propagation in the single crystal. However, this beam gets the
circular polarization only in the case when $X_2 \ne 0$ (i.e., normal waves
are elliptically polarized). Taking into account Eq.(9), one can expect
that the circular polarization may be appreciable only on a significant 
thickness.

The value of intensity of $\gamma$-quanta on any thickness x depends on their
initial polarization. It is convenient to define the following 
value of asymmetry:
\begin{eqnarray} 
A_P(x|P)= {{J_\gamma(x|P)-J_\gamma(x|0,0,0)} \over J_\gamma(x|0,0,0)},
\end{eqnarray}
where the matrix $P = (\xi_1(0), \xi_2(0), \xi_3(0))$. It easy to see that
the $A_P$-asymmetry is equal to the relative variation of intensity between
the two cases; in one case the $\gamma$-quanta is initially unpolarized,
and, in second case the $\gamma$-quanta have the polarization, 
which  described by the matrix P. Any asymmetry $A_P$ can be expressed with the
use of the three basic asymmetries $ A_1, A_2$ and $A_3$ in the following form:
\begin{eqnarray}
A_P(x)= A_1(x)\xi_1(0)+A_2(x)\xi_2(0)+A_3(x)\xi_3(0).
\end{eqnarray}
The basic asymmetries $A_1(x)$, $A_2(x)$ and $A_3(x)$ are determined
correspondingly by the next P-matrices: (1,0,0), (0,1,0), (0,0,1). 

Besides, the $A_i(x)$-parameters  connect with the corresponding
Stokes parameters ${\cal{P}}_i(x)\,$ (i=1-3) for initially unpolarized 
$\gamma$-quanta on the same thickness by the following relations 
\begin{eqnarray}{\cal{P}}_1(x)=A_1(x), \, {\cal{P}}_2(x)=-A_2(x), \,
 {\cal{P}}_3(x)=A_3(x).
\end{eqnarray} 
Note that the positive quantity of $\xi_2$ corresponds to the 
right circular polarization.

\section{Calculations and discussion}
We have considered the process of 50-GeV $\gamma$-quanta propagation in
the silicon single crystal at room temperature. The case, when the beam of 
$\gamma$-quanta
move under small angle (at some milliradians) with respect to $<001>$ axis
of the single crystal, was selected for computations. 
  In calculations the Moliere atomic form factor was employed \cite{Mo}.

One can determine the direction of $\gamma$-quanta motion with the use
of  angle $\theta$ with respect to $<001>$-axis and the azimuth 
angle $\alpha$ around of this axis ($\alpha =0$ when the momentum of 
$\gamma$-quanta lies in  the  $(110)$  plane).  However,  another  angles  
are  more convenient to
use $\theta_H= \theta \cos\alpha, \theta_V= \theta\sin\alpha$.  

Fig.1 illustrates the calculations of the $|X_2|$-value as a function of 
the two angles $\theta_H$ and $\theta_V$. 
The numbers on Fig.1 show the degree of circular polarization of normal waves. 
The number 0 corresponds to $|X_2|$-values  from 0 to 0.1, number 1
corresponds to ones from 0.1 to 0.2 and so on. One can see that
the widths of both orientation angles where $|X_2|>0.5$ 
is about 0.3-0.4 mrad. 
The same picture takes a place also at different energies of $\gamma$-quanta.   
However,  in this case the angles of orientation  depend inversely proportional 
on the energy.

 Fig.2 illustrates  the  calculations  of  the  absolute  values  
of basic asymmetry parameters
for the two orientations of the single crystal as a function of 
the thickness $x$. 
One can see that  $|A_2(x)|<< |A_3(x)|$ if thickness $x< 10-15$ cm.  
Note that intensity of initially unpolarized $\gamma$-quanta is equal to
0.0082 at x=25 cm (0.000078 at x=50 cm) for the first orientation (solid lines) 
and correspondingly  0.0058 (0.000034) for the second orientation
(dashed lines). The $|X_2|$-values are equal to 0.88 and 0.48 for these
orientations. Note, we use in calculations the Cartesian coordinate
system with one axis along the momentum of $\gamma$-quanta and with the
other two axes lying in the $(110)$ and $(1 \overline{1} 0)$ crystallographic 
planes. 

     The polarization state of propagating $\gamma$-beam become same as the
state of one normal wave begining from some thickness. However, 
it takes a place
at the very large values of $x > 500$ cm (for curves on fig.2), 
and it is out of significance to the practicality.  The reason is the small
value of $\tilde n_1 - \tilde n_2$ for high circular polarizations
of normal waves 
( $\tilde n_1 - \tilde n_2 \rightarrow 0 \, \,at \, \, X_2 \rightarrow \pm 1$).    

  The sensitivity of single crystals to the circular polarization
can be used for its measurements. In the general case initial beam of 
$\gamma$-quanta has the linear and circular polarizations. As  follows from
Eq.(22) two measurements of asymmetry $A_P(x)$  make possible
determination of the circular polarization of $\gamma$-quanta. 
Let $A_P(x|\xi_1(0), \xi_2(0), \xi_3(0))$ be the measured asymmetry in the
first measurement. In the second measurement, it needs to change  
the direction of linear polarization on $90^o$ with respect to the axis of
single crystal (the $\theta_H, \theta_V$ are the same in both cases).
It is possible to make by rotating of the single crystal around the 
momentum of $\gamma$-quanta. As a result we get 
\begin{eqnarray} 
A_P((x|\xi_1(0),\xi_2(0), \xi_3(0))+ A_P(x|-\xi_1(0),\xi_2(0),-\xi_3(0)) 
= 2A_2(x)\xi_2(0).
\end{eqnarray}   
One can rewrite this equation in the following form:
\begin{eqnarray}J_\gamma (x|\xi_1(0), \xi_2(0), \xi_3(0))
+ J_\gamma (x|-\xi_1(0), \xi_2(0), -\xi_3(0)) =
\nonumber \\  =2J_\gamma(x|0,0,0)+  2 J_\gamma(x|0,0,0) A_2(x) \xi_2(0). 
\end{eqnarray} 
Knowing the $J_\gamma(x|0,0,0)$ and $A_2(x)$-values , 
one can determine the circular
polarization of $\gamma$-quanta if the relative intensities 
$J_\gamma(x|\pm \xi_1(0) ,\xi_2(0), \pm \xi_3(0))$ were found experimentally. 
\section{Conclusion}
We have considered the general case of propagation of $\gamma$-quanta in
single crystals. It was shown that there exist some regions of orientations
where single crystals are sensitive to circular polarization of $\gamma$-beam,
and the possibility of its measurements is discussed.  
Note that there are few in number of processes in the condensed medium 
sensitive to the circular polarization of $\gamma$-quanta with energies 
in tens GeV.   

The obtained results may be useful for creation of polarimeters for high
energy beams of $\gamma$-quanta.



\newpage
\begin{figure}
[h]\begin{center}
\parbox[c]{13.5cm}{\epsfig{file=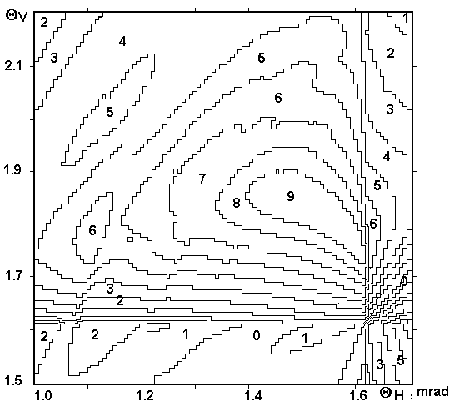,height=12cm}}
\parbox[c]{15cm}{\caption{
 The $|X_2|$-value as a function of orientation angles $\theta_H$ and
$\theta_V$ for $50-GeV$-$\gamma$-quanta in the silicon single 
crystal. Further explanations are given in the text.
              }}  
\end{center} 
\end{figure}
\newpage
\begin{figure}[h]
\begin{center}
\parbox[c]{13.5cm}{\epsfig{file=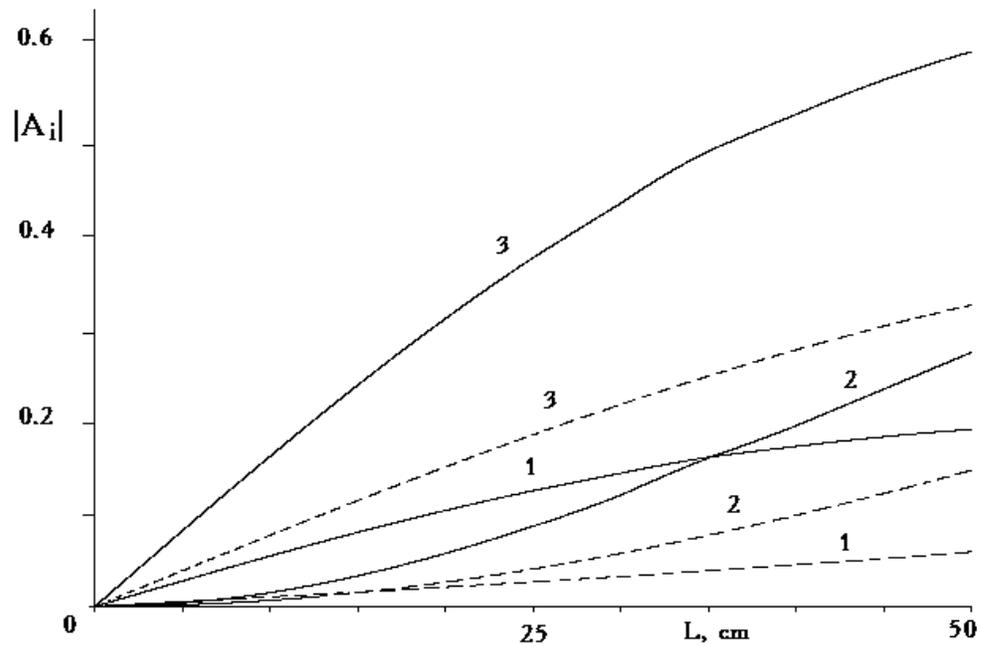,height=9cm}}
\parbox[c]{15cm}{\caption{
The absolute values of basic 
asymmetries $A_i,\,(i=1-3)$ as functions of the single crystal thickness.
The calculated values of asymmetries for angles $\theta_H = 1.6\,mrad, 
\, \theta_V= 1.8 \, mrad$ are presented by the solids curves, and for
$\theta_H=1.0 \, mrad, \, \theta_V = 1.8 \, mrad$ - by the dashed curves.
             }}  
\end{center} 
\end{figure}
}
\end{document}